\documentclass[pra,dvips,twocolumn]{revtex4}
\usepackage{epsfig,graphics,graphicx,psfig}
\begin{document}
\title{Squeezing generation and revivals in a cavity-ion system in contact with a reservoir}
\author{R. Rangel}
 \email{renata@if.ufrj.br}
\author{L. Carvalho}
\author{N. Zagury}%
\affiliation{Instituto de F\'\i sica, Universidade Federal do Rio de Janeiro,Caixa 
Postal 68528, 21945-970, Rio de Janeiro, Brazil
}%

\date{today}
\begin{abstract}
We consider a system consisting of a single two-level ion in a harmonic trap, which is localized inside a non-ideal optical cavity at zero temperature and subjected to the action of two external lasers. We are able to obtain an analytical solution for the total density operator of the system and show that squeezing in the motion of the ion and in the cavity field is generated. We also show that complete revivals of the states of the motion of the ion and of the cavity field occur periodically. 
\end{abstract}
\maketitle
\vskip2pc

\section{Introduction}

In the last two decades much attention has been given to the study of squeezing both theoretically and experimentally.
Squeezed states may be used to improve the signal-to-noise ratio in optical communications \cite{yuen} and in spectroscopic experiments \cite{wineland1,wineland2}. There is hope that it can also be used in very sensitive experiments, like the detection of gravitational waves \cite{hollenhorst,caves}.

Research on squeezing has been done in several contexts. Here we are particularly interested in systems of atoms or ions inside an electromagnetic cavity. It is well known that squeezing may be obtained in the interaction of one mode of the cavity with a two-level or a three-level atom. The one photon and the two photon micromasers \cite{micromaser} and microlasers \cite{an} are examples of such systems. Selective atomic measurements in cavity Q.E.D. have been proposed to enhance squeezing (up to $75\%$) in the Jaynes-Cummings model \cite{gerry}.
Villas-B\^oas {\it et al}. \cite{villasboas} proposed squeezing an arbitrary radiation field state previously prepared in a high-Q cavity by the dispersive interaction of a single three-level atom with a classical field and a cavity mode. 
Massoni and Orszag \cite{massonisq} introduced a novel way to produce squeezed light in an optical cavity through the transferring  the squeezing from the motion of a three-level trapped ion to the cavity mode. Two external laser fields drive the atom and the squeezing in the motion is generated by an external electric field.    
Here we have considered the simpler case of a two-level ion trapped inside the cavity.  
This system was originally considered by Zeng and Lin \cite{zeng} and developed by many authors \cite{system}. 
In Ref.~\cite{rangel} an analytical solution of the master equation was obtained in the case that the ion is inside a bad cavity  and is excited by only one laser field. In this case the energy may be transfered back and forth from the cavity field to the motion of the ion. 

In this paper we discuss a scheme for generating squeezing both in the motion of an ion, driven by two external laser fields, and in a mode of the electromagnetic field inside a non-ideal cavity at zero temperature. Even though we consider a non-ideal cavity, we are able to obtain an {\it analytical} solution for the total density operator at any time and show that the motion and the cavity field are {\it always squeezed} in one of their quadratures if the ratio between the intensity of the lower and higher frequency fields is less than one. Remarkably this system presents revivals, the vibrational state periodically turns into the {\it same squeezed vacuum state and the cavity field periodically visits the vacuum state}. Although we are concerned here to a specific system, our analytical results may be useful in any situations were the effective Hamiltonian corresponds to a general linear coupling of two oscillators, one of them being connected to a reservoir at zero temperature. In Sec.~II we present the model that couples the ion to the cavity field. In Sec.~III we give an analytical solution for the master equation showing explicitly the revivals and study the generation of squeezing. In Sec.~IV we summarize our results and present our conclusions. In the Appendix we present a detail calculation for obtaining the solution presented in Sec.~III.    

\section{The model}

The system we are considering is very similar to the one discussed in Ref.~\cite{rangel}. A two-level ion is bounded by a linear Paul trap, which is inside an optical cavity. The ion has mass $m$ and the effective potential created by the Paul trap may be approximated by $(1/2)m\nu^2 \hat x^2,$ where $\hat x$ is the displacement operator of the ion from its equilibrium position and $\nu$ is the classical frequency of oscillation of a particle in this potential. The trapping direction $x$ coincides with the axis of the cavity. The electronic levels of the ion, $\vert e \rangle$ and $\vert g \rangle,$ are separated by the energy $\hbar \omega_0$ and are quasi resonant to a stationary mode of the cavity of frequency $\omega_c.$ This system is then described by the Hamiltonian $\hat H:$ 
\begin{eqnarray}\label{H0}
\hat H&=&\hat H_0+\hat H_1\, ,  \nonumber \\
\hat H_0 &=& \hbar\omega_0 \vert e \rangle\langle e\vert + \hbar\omega_c\hat a^{\dagger}\hat a + \hbar\nu \hat b^{\dagger}\hat b \, , \nonumber\\
\hat H_1 &=& \hbar  g_c \sin( k_c\, \hat x ) \, \hat a \, \vert e \rangle \langle g \vert+{\rm h. c.}  \, ,
\end{eqnarray}
where $\hat a$ and $\hat a^{\dagger}$ ($\hat b $ and $\hat b^{\dagger}$) are the annihilation and creation operators of the cavity mode (vibrational quanta).
In $\hat H_1,$ $k_c$ is the cavity wavevector and $g_c$ is  the ion-cavity  coupling constant. We have taken a sinus function as the cavity standing wave mode and set the minimum of the trapping potential at a cavity node.
 The eigenstates of $\hat H_0$  are tensor products of the electronic states,  $\vert e \rangle$ and $\vert g \rangle,$ times  Fock states $\vert n \rangle_c$ and $\vert m \rangle_v,$ associated to the cavity field and vibrational quanta. The position operator is related to the operators $\hat b$ and $\hat b^{\dagger}$ by $\hat x = \delta x (\hat b + \hat b^{\dagger}),$  with $\delta x = \sqrt{\hbar / (2m\nu)}$ being the uncertainty of position in the vibrational ground state. 

Now we let two external lasers act on the ions. The laser frequencies, $\omega_1 = \omega_c - \nu $  and $\omega_2 = \omega_c + \nu,$ and the detuning, $\Delta= \omega_0 - \omega_c,$ are chosen in order that Raman transitions among the 
level $\vert g  \rangle \vert n \rangle_c \vert m \rangle_v$ and the levels  $\vert g  \rangle \vert n+ 1 \rangle _c  \vert m \pm 1 \rangle_v , $ $\vert g  \rangle \vert n- 1 \rangle _c  \vert m \pm 1 \rangle_v$ may occur (see Fig.~\ref{Figura1}). The total interaction Hamiltonian that describes the coupling of the internal and external degrees of the ion with the cavity and with the laser fields, in the interaction picture with respect to $\hat H_0,$ may be written as  
\begin{eqnarray}\label{Hint} 
\hat H_{\rm int}(t) &=& \hbar\{ g_c \sin( k_c \hat x(t) ) \, \hat a \, e^{i\Delta t} 
- i g_1 e^{ i k_1 \hat x(t)+ i(\Delta+\nu) t} \nonumber \\ 
&&- i g_2  e^{ i k_2 \hat x(t) +i (\Delta -\nu)t} \}\vert e \rangle \langle g \vert+{\rm h. c.},
\end{eqnarray}
where $k_1, k_2$ are the $x$ components of the laser wavevectors and  $g_1, g_2$ are the  ion-laser coupling constants.  $\hat x(t)$ is the position operator in the interaction picture.

\begin{figure}
\resizebox{!}{4cm}{\includegraphics{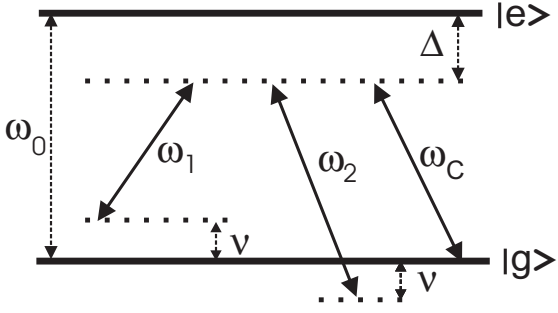}}
\caption{\label{fig:Figura1} Level scheme of the ion. The excited level $\vert e \rangle$ and the ground level $\vert g \rangle$ are coupled by laser fields of frequencies $\omega_1$ and $\omega_2$ and one cavity mode of frequency $\omega_c,$ detuned by $\Delta.$ $\omega_0$ and $\nu$ are the atomic transition and the vibration frequencies, respectively.}
\end{figure}

We consider that the Lamb-Dicke limit $(\eta_c =  k_c\delta x \ll 1, $ $\eta_1 = k_1\delta x \ll 1,$  $\eta_2 = k_2\delta x \ll 1 )$ is valid. We also assume the rotating wave approximation and that $\omega_0\gg \Delta\gg \eta_1 g_1, \eta_2 g_2, \eta_c g_c  ,\nu . $ For $t\gg 1/\Delta,$ we obtain, following the usual procedure for the adiabatic elimination of one of the electronic levels\cite{luiz}, the effective interaction Hamiltonian:
\begin{eqnarray}\label{V}
\hat V_{\rm eff}^\prime & = &   
- i \hbar \Omega_1 \hskip 0.1cm (\hat a^{\dagger}\hat b - \hat a \hat b^{\dagger})\hat \sigma_z - i  \hbar \Omega_2 \hskip 0.1cm (\hat a^{\dagger}\hat  b^{\dagger} - \hat a\hat b)\hat \sigma_z  \nonumber \\
&+& \hbar( g_1^2 + g_2^2)/ \Delta \, \hat \sigma_z \, ,
\end{eqnarray}
where $\hat \sigma_z = \vert e \rangle \langle e \vert - \vert g \rangle \langle g \vert,$
$\Omega_1 = \eta_c g_1 g_c / \Delta$ and \linebreak $\Omega_2 =  \eta_c g_2 g_c/ \Delta.$ The third term is the usual Stark shift and will be incorporated in $\hat H_0$ by redefining $\omega_0.$ Non linear terms in $\eta$ will be neglected by assuming that the average number of excitations are not very large.  If the ion starts in its ground state, we may substitute the operator $\hat \sigma_z$ by $-1.$ In this case, the effective Hamiltonian (\ref{V}) may be rewritten as
\begin{equation}\label{Veff}
\hat V_{\rm eff} = i \hbar \Omega_1 \hskip 0.1cm (\hat a^{\dagger}\hat b - \hat a\hat b^{\dagger}) +i  \hbar \Omega_2 \hskip 0.1cm (\hat a^{\dagger}\hat  b^{\dagger} - \hat a\hat b ) \, .
\end{equation}
The first term of Eq.~(\ref{Veff}) describes processes where the sum of the number of photons and of vibrational quanta are conserved while the second term describes processes where the same number of photons and vibrational quanta are created or absorbed simultaneously. These processes allow the exchange of quantum information between the cavity field and the vibrational motion and  the generation of squeezing. 

\section{The density operator}

Sources of dissipation in the mechanical movement inside a Paul trap may be attributed to chaotic fields due to strain fields in the trap electrodes and to the electronic decay of the upper level $|e\, \rangle$~\cite{vogel, adrian}. They are much easier to control than the losses of the cavity field. For example, heating in recent Paul trap experiments have been limited to $1$ quantum per $6$ms~\cite{rowe}. Also the heating due to the electronic decay could be made small if the detuning is sufficiently large. Here we only consider cavity losses and neglect the interaction of the ion vibrational motion with the environment. Assuming a linear interaction of the cavity field with a reservoir at zero temperature and using the Born-Markov approximation we may write a master equation were the losses are included via a Lindblad form \cite{Cohen}  
\begin{equation}\label{eqm}
\frac {\partial \hat \rho_T(t)} {\partial t} = \frac {1}{i \hbar} [\hat V_{\rm eff}, \hat \rho_T (t)] + {\cal L} \hat \rho_T (t)\, ,
\end{equation}  
where $\hat V_{\rm eff}$ is the effective Hamiltonian (\ref{Veff}) and
\begin{equation}\label{L}
{\cal L} \hat \rho_T = \frac {\gamma}{2}(2 \hat a \hat \rho_T \hat a^{\dagger} - \hat a^{\dagger} \hat a \hat \rho_T - \hat \rho_T \hat a^{\dagger} \hat a) \, .
\end{equation}

In the Appendix, we solve this master equation when the initial state is the vacuum state, \linebreak $\hat \rho_T(0) = \vert 0 \rangle_c \hskip 0.1cm _c{\langle} 0 \vert \hskip 0.1cm \vert 0 \rangle_v  \hskip 0.1cm  _v{\langle} 0 \vert \, . $ We obtain for the density operator at time $t$ 
\begin{eqnarray}\label{Rhot}
\hat \rho_T(t) &=& \sum_{m=0}^{\infty} \hskip 0.1cm \sum_{n=0}^{\infty} \hskip 0.1cm (f(t)g(t))^{m+n} \nonumber \\
&\times& \hat Q^{m,n}_{\hskip 0.1cm c}(\bar n_c(t),\xi_c(t))  \hskip 0.1cm \hat Q^{m,n}_{\hskip 0.1cm v}(\bar n_v(t),\xi_v(t))\, ,
\end{eqnarray}
where the functions $f(t)$ and $g(t)$ are given by
\begin{eqnarray}\label{ftgt}
f(t) &=& \biggl( \cos(\Lambda t) + \frac {\gamma} {4 \Lambda}\sin(\Lambda t) \hskip 0.1cm \biggr) \hskip 0.1cm   e^{- \gamma t/4} \, , \nonumber \\
g(t) &=& \frac {\Omega_2} {\Lambda }\sin ( \Lambda t) \hskip 0.1cm e^{- \gamma t/4} \, ,
\end{eqnarray}
with $\Lambda^2 = \Omega_1^2 - \Omega_2^2 - \gamma^2/16.$
The operators $\hat Q^{m,n}_{\hskip 0.1cm \sigma}(\bar n,\xi),$ with $\sigma = c,v,$ labeling the cavity field and vibrational motion respectively, are defined as 
\begin{eqnarray}\label{Qmnnxi}
&&\hat Q^{m,n}_{\hskip 0.1cm \sigma}(\bar n,\xi)  \nonumber \\
&& = \sum_{k=0}^{m+n} C^{m,n}_{\hskip 0.1cm k}(\xi) \, \hat S_{\sigma}(\xi) \, \hat R^{m+n-k,k}_{\hskip 0.2cm \sigma} (\bar n) \, \hat S_{\sigma}(\xi)^{\dagger} \, ,
\end{eqnarray}
with $\hat S_{\sigma}(\xi)$ being the  squeezing operators that act on the cavity or in the vibrational states: 
\begin{eqnarray}\label{S}
\hat S_c(\xi) &=& e^{\xi/2 (\hat a^2 - \hat a^{\dagger \hskip 0.05cm 2})} \, , \nonumber \\ 
\hat S_v(\xi) &=& e^{\xi/2 (\hat b^2 - \hat b^{\dagger \hskip 0.05cm 2})} \, .
\end{eqnarray}
The coefficients $C^{m,n}_{\hskip 0.1cm k}(\xi)$ are given by
\begin{eqnarray}\label{Cmn}
&&C^{m,n}_{\hskip 0.1cm k}(\xi) = \sqrt{\frac{(m+n-k)!k!}{m!n!}} \nonumber \\
&\times& \sum_{l={\rm max}(0,k-m)}^{{\rm min}(n,k)} \frac{m!}{(k-l)!(m-k+l)!}\frac{n!}{l!(n-l)!} \nonumber \\
&\times& (\cosh \xi)^{m-k+2l} (\sinh \xi)^{n+k-2l}
\end{eqnarray}
and the operators $\hat R^{m,n}_{\hskip 0.1cm \sigma},$ which satisfy the relation $ \hat R_{\hskip 0.1cm \sigma}^{n,m}=( \hat R_{\hskip 0.1cm \sigma}^{m,n} )^{\dagger},$ are defined for $m \geq n$ as
\begin{eqnarray}\label{Rmnn}
&&\hat R^{m,n}_{\hskip 0.1cm \sigma}(\bar n)  = \sum_{k = 0}^{\infty} \hskip 0.1cm \sqrt{\frac{n! k! }{ m! (k + m -n)!}} \hskip 0.1cm \frac{1 }{ (\bar n + 1)^{m+1}}  \nonumber \\
&\times& P_{\hskip 0.1cm \hskip 0.1cm m}^{\hskip 0.1cm k,k -n} \biggl(\displaystyle{\frac{\bar n }{ \bar n +1}}\biggr) \vert k + m -n \rangle_{\sigma} \, _{\sigma}\langle k\vert \, , 
\end{eqnarray}
where 
\begin{equation}\label{jacoby}
P_{\hskip 0.1cm \hskip 0.1cm m}^{\hskip 0.1cm k,l}(x) = \sum_{j={\rm max} (0,l)}^{k} (-1)^{j-l} \frac{(j + m)! }{ (j-l)! (k-j)!} \frac{x^j }{ j!}\, ,
\end{equation}
may be identified as a Jacoby polynomial. The functions  $\bar n_{c}(t),$ $\bar n_{v}(t),$  $\xi_{c}(t)  $  and $\xi_{v}(t)  $ in Eq.~(\ref{Rhot}) are given by 
\begin{eqnarray}\label{zetaxin}
\bar n_{\sigma}(t) &=& - 1/2 + \sqrt{1/4+\nu_{\sigma}(t)-(q^2-1)\,  \nu_{\sigma}(t)^2 }  \, , \nonumber \\
\vert \xi_{\sigma}(t) \vert &=& \frac{1}{4} \hskip 0.1cm {\rm ln} \bigg( \frac{1/2 + (q+1)\, \nu_{\sigma}(t)  }{1/2 - (q-1) \, \nu_{\sigma}(t)} \bigg) \, , \hskip 0.5cm  \sigma=c,v, \nonumber \\
\xi_{c}(t) &=& - \vert \xi_{c}(t) \vert \hskip 0.5cm , \hskip 0.5cm \xi_{v}(t) = \vert \xi_{v}(t) \vert \, ,
\end{eqnarray}
where $q=\Omega_1/\Omega_2, $  $\nu_c(t)=  g(t)^2 \,$ and
\begin{equation}\label{munut}  
 \nu_v(t)=\frac{1}{q^2 - 1} \hskip 0.1cm (1 - f(t)^2) \, .
\end{equation}

The reduced density operators are obtained by taking the partial traces of Eq.~(\ref{Rhot}). By using the property (see Eq.~(\ref{trQmnA})),
\begin{equation}\label{trQmn}
{\rm tr}_{\sigma} \hat Q^{m,n}_{\hskip 0.1cm \sigma}(\bar n, \xi) = \delta_{m,0} \delta_{n,0} \, ,
\end{equation}
we easily get
\begin{equation}\label{rhosigma}
\hat \rho_{\sigma}(t) = \hat Q^{0,0}_{\hskip 0.1cm \sigma}(\bar n_{\sigma}(t),\xi_{\sigma}(t)) \quad , \quad  \sigma = c,v \, .
\end{equation}
In the Appendix we show that $ \hat Q^{0,0}_{\hskip 0.1cm \sigma}(\bar n,\xi)$ has the form of a ``squeezed thermal state" (see Eq.~(\ref{Q00nxiA})):
\begin{eqnarray}\label{Q00nxi}
\hat Q^{0,0}_{\hskip 0.1cm \sigma}(\bar n,\xi)= \sum_{k=0}^{\infty} \frac{\bar n^k}{(\bar n+1)^{k+1}} \hat S_{\sigma}(\xi) \, \vert k \rangle_{\sigma} \, _{\sigma}\langle k \vert \, \hat S_{\sigma}(\xi)^{\dagger} \, . 
\end{eqnarray}

\subsection{ $\bf{\Omega_1^2>\Omega_2^2:}$ squeezing generation and revivals}

When $\Omega_1^2>\Omega_2^2, $ the system reaches a steady state. From Eqs.~(\ref{ftgt}),(\ref{zetaxin}),(\ref{munut}), we see that in the limit $t \rightarrow \infty, $ $\bar n_v(t), \bar n_c(t),\xi_c(t)\rightarrow 0 .$  Therefore the total density operator in the steady state may be written as 
\begin{equation}\label{steady}
\hat \rho_T^{\, \rm st}= \vert 0\rangle_c \,_c\langle 0 \vert \, \hat \rho^{\, \rm st}_v \, ,
\end{equation}
with
\begin{equation}\label{rhov}
\hat \rho^{\, \rm st}_v=
\hat S_v(\bar\xi) \hskip 0.1cm \vert 0 \rangle_v \hskip 0.1cm _v\langle 0 \vert \hskip 0.1cm \hat S_v(\bar\xi)^{\dagger} \, ,
\end{equation}
where 
\begin{equation}\label{xi}
\bar\xi = \frac{1}{2} {\rm ln} \bigg( \frac{\Omega_1 + \Omega_2}{\vert\Omega_1 - \Omega_2\vert} \bigg) \, .
\end{equation}
Eq.~(\ref{steady}) shows that when the system reaches the steady state the energy transferred from the lasers to the cavity dissipates away while the vibrational state becomes a squeezed vacuum state. 

When $\Omega_1^2 > \Omega_2^2 + \gamma^2/16$ the function $f(t)$ vanishes at times
\begin{equation}\label{taun}
\tau_n = \frac{1}{\Lambda} \arccos \pmatrix{\displaystyle{\frac{-\gamma/4}{\sqrt{\Omega_1^2 - \Omega_2^2}}}} + n \frac{\pi}{\Lambda} \, ,
\end{equation}
with $\pi/2< \arccos x <\pi$ and $n=0,1,2,\dots.$ Since $f(\tau_n)=0,$ the cavity field and the ion motion are instantly decorrelated at $t=\tau_n:$  
\begin{equation}\label{zeta=0v}
\hat \rho_T(\tau_n) = \hat \rho_c(\tau_n) \hat \rho_v(\tau_n)\, ,
\end{equation}
as can be seen from Eq.~(\ref{Rhot}). Also at these times $\bar n_v(\tau_n)=0$ and $\xi_v(\tau_n)=\bar\xi.$  From Eq.~(\ref{Q00nxi}) we obtain 
\begin{equation}\label{Rhotn}
\hat \rho_v(\tau_n)= \hat S_v(\bar\xi) \hskip 0.1cm \vert 0 \rangle_v \hskip 0.1cm _v\langle 0 \vert \hskip 0.1cm \hat S_v(\bar\xi)^{\dagger} = \hat \rho^{\, \rm st}_v,
\end{equation}
that is, the reduced operator that describes the vibrational system returns to the same state periodically and this state coincides with the reduced operator in the steady state. Notice that for these finite times the cavity state remains a ``squeezed thermal state" with $\bar n_c(\tau_n)$ varying as $\tau_n$ increases. 

There is also recurrences in the state of the cavity field at times 
\begin{equation}\label{taunprime}
\tau_n^\prime = n \frac{\pi}{\Lambda}  \hskip 0.4cm , \hskip 0.4cm n=0,1,2,\dots \, , 
\end{equation}
when $\xi_c(\tau_n^\prime)=\bar n_c(\tau^{\prime}_n)=0,$ since $g(\tau_n^\prime)=0.$ Therefore the cavity field and the motion are also instantly decorrelated at times $\tau_n^\prime,$
\begin{equation}\label{zeta=0c}
\hat \rho_T(\tau_n^\prime) = \hat \rho_c(\tau_n^\prime) \hat \rho_v(\tau_n^\prime)\, ,
\end{equation}
and the cavity field returns, with a period $\pi/\Lambda,$ to the vacuum state, 
\begin{equation}\label{Rhotnprime}
\hat \rho_c(\tau_n^\prime) = \vert 0 \rangle_c \hskip 0.1cm _c\langle 0 \vert \, .
\end{equation}
For small values of $\Omega_2/\Omega_1, $ the ``squeezed thermal state" given in Eq.~(\ref{Q00nxi}),
is very close to an ideal squeezed state. In fact, the maximum value for $\bar n_c(t)$ and $\bar n_v(t)$ is given by 
\begin{equation}
\bar n_{\rm max}=\frac 12\bigg( \frac{1}{\sqrt{1-(\Omega_2/\Omega_1)^2}}-1\bigg)\, ,
\end{equation}
which is approximately $ 0.25(\Omega_2/\Omega_1)^2,$  for $(\Omega_2/\Omega_1)^2\ll 1.$

Having obtained the solution in the case that the initial state is the vacuum, Eq.~(\ref{Rhot}), we  may easily find analytical solutions of the master equation when the initial state is the product of coherent states:
\begin{equation}\label{rho0alphabeta}
\hat \rho_T^{\alpha,\beta}(0) = \hat D_{c}(\alpha)\vert 0 \rangle_{c} \hskip 0.1cm _{c}\langle 0 \vert \hat D_{c}(\alpha)^\dagger \hskip 0.1cm \hat D_{v}(\beta) \vert 0 \rangle_{v} \hskip 0.1cm _{v}\langle 0 \vert \hat D_{v}(\beta)^\dagger \, , 
\end{equation}
where 
\begin{eqnarray}\label{DcDv}
&&\hat D_{c}(\alpha) = e^{\alpha \hat a^{\dagger} - \alpha^* \hat a} \, , \nonumber \\
&&\hat D_{v}(\beta) = e^{\beta \hat b^{\dagger} - \beta^* \hat b} \, .
\end{eqnarray}
In this case we have  
\begin{equation}\label{rhotalphabeta}
\hat \rho_T^{\alpha,\beta}(t) = \hat D_{c}(u(t)) \, \hat D_{v}(v(t))   \hskip 0.1cm \hat \rho_T(t)  \hskip 0.1cm \hat D_{v}(v(t))^{\dagger} \, \hat D_{c}(u(t))^{\dagger}\, ,
\end{equation}where $\hat \rho_T(t)$ is the density operator for the initial vacuum state, which is  given in Eq.~(\ref{Rhot}), and $u(t)$ and $v(t)$ are defined as 
\begin{eqnarray}\label{uev}
&&u(t) = \alpha \, h(t) + (\beta \Omega_1/ \Omega_2 + \beta^* )g(t) \, , \nonumber \\
&&v(t) = (- \alpha \Omega_1/ \Omega_2 + \alpha^* )g(t) + \beta f(t)\, .
\end{eqnarray}
The functions $f(t)$ and $g(t)$ were already given in Eq.~(\ref{ftgt}) and $h(t)$ is defined as 
\begin{equation}\label{ht}
h(t)=\bigg(\cos \Lambda t - \frac{\gamma}{4 \Lambda} \sin \Lambda t \bigg)e^{-\gamma t/4}\, .  
\end{equation}
That is, the solution of the master equation, when the initial state is the vacuum displaced by $\alpha$ and $\beta$ in the phase space of the whole system, corresponds to a displacement of $u(t)$ and $v(t)$ of the solution of the master equation when the initial state is the  vacuum. Then the  reduced density operators may be  written as
\begin{eqnarray}\label{Rhoabcv}
\hat \rho^{\alpha,\beta}_{c}(t)&=& \hat D_{c}(u(t))\hskip0.1cm 
\hat \rho_c(t)\hskip0.1cm  \hat D_{c}(u(t))^{\dagger}\, , \nonumber\\
\hat \rho^{\alpha,\beta}_{v}(t)&=& \hat D_{v}(v(t))\hskip0.1cm 
\hat \rho_v(t)\hskip0.1cm  \hat D_{v}(v(t))^{\dagger}\, ,
\end{eqnarray}
where $\hat \rho_c(t)$ and $\hat \rho_v(t)$ are given in Eq.~(\ref{rhosigma}). 
When $\Omega_1^2 > \Omega_2^2$, for any values of $\alpha $ and $\beta,$ the cavity field and the motion of the ion are decorrelated instantly at  times $\tau_n$ and $\tau_n^\prime, $ given in Eqs.~(\ref{taun}) and~(\ref{taunprime}). For  $\alpha=0$ recurrences  will also occur at these times, since when $t= \tau_n,$ $f(\tau_n)=0  $  and when $t= \tau_n^\prime,$ $g(\tau_n^\prime)=0 . $ As a consequence $v(\tau_n),$ $\bar n_v(\tau_n),$  $u(\tau_n^\prime),$ $\bar n_c(\tau_n^\prime)$ vanishes and $\xi_v(\tau_n)=\bar\xi, $ $\xi_c(\tau_n^\prime)=0. $ Therefore  these revivals occur, independently of the value of $\beta, $ at $t=\tau_n$ to  the {\it same} squeezed vacuum vibrational state obtained before and  given by  Eq.~(\ref{Rhotn}), and at $t=\tau_n^\prime$ to the cavity vacuum, Eq.~(\ref{Rhotnprime}). In fact, we have verified  that this last result is valid for {\it any} initial state of the motion as long as the initial state of the cavity field is the vacuum. 

 Using Eqs.~(\ref{rhosigma}), (\ref{Q00nxi}), (\ref{Rhoabcv}), we may show that the uncertainties associated to the quadratures
\begin{eqnarray}\label{PcXv}
\hat X_c =  \frac{ \hat a + \hat a^{\dagger}}{\sqrt{2}} \quad , \quad \hat P_c  &=&  \frac{ \hat a - \hat a^{\dagger}}{i\sqrt{2}} \, ,\nonumber \\
\hat X_v =  \frac{\hat b + \hat b^{\dagger}}{\sqrt{2}}  \quad , \quad \hat P_v  &=&  \frac{ \hat b - \hat b^{\dagger}}{i\sqrt{2}} \, ,
\end{eqnarray}
have the form
\begin{eqnarray}\label{DXDP}
\Delta X_{\sigma} (t)^2 &=& ( \, \bar n_{\sigma}(t) + 1/2 \, ) \, e^{-2 \xi_{\sigma}(t)} \, , \nonumber \\
\Delta P_{\sigma} (t)^2&=& ( \, \bar n_{\sigma}(t) + 1/2 \, ) \, e^{2 \xi_{\sigma}(t)} \, , \hskip0.5cm \sigma =c,v\, ,
\end{eqnarray}
in the case that the initial state is a coherent state. Notice  that  $\Delta X_{\sigma} (t)^2 \Delta P_{\sigma} (t)^2=( \, \bar n_{\sigma}(t) + 1/2 \, )^2.$

By substituting the expressions of $\bar n_{\sigma}(t)$ and $\xi_{\sigma}(t)$ into the above equations, it is easy to show that:
\begin{eqnarray}\label{DXDPfg}
\Delta X_c (t)^2 &=& \frac{1}{2} + \frac{\Omega_1 + \Omega_2}{\Omega_2} g(t)^2   \, ,\nonumber \\
\Delta P_c (t)^2 &=&  \frac{1}{2} - \frac{\Omega_1 - \Omega_2}{\Omega_2} g(t)^2 \, , \nonumber \\
\Delta X_v (t)^2 &=& \frac{1}{2} - \frac{\Omega_2}{\Omega_1 + \Omega_2}(1 - f(t)^2) \, ,\nonumber \\
\Delta P_v (t)^2 &=&\frac{1}{2}+\frac{\Omega_2}{\Omega_1 - \Omega_2}(1 - f(t)^2) \, ,
\end{eqnarray}
that is, when $\Omega_1>\Omega_2,$ the quadrature $\hat X_v$ is always squeezed for $t> 0,$ while the quadrature $\hat P_c$ is equal to $1/2$ when  $t=\tau_n^\prime=n\pi/\Lambda $ and is squeezed otherwise, since $(1-f(t)^2)>0$ for $t > 0$ and $g(t)^2>0$ for $t\neq \tau_n^\prime\, .$

\begin{figure}
\resizebox{!}{12cm}{\includegraphics{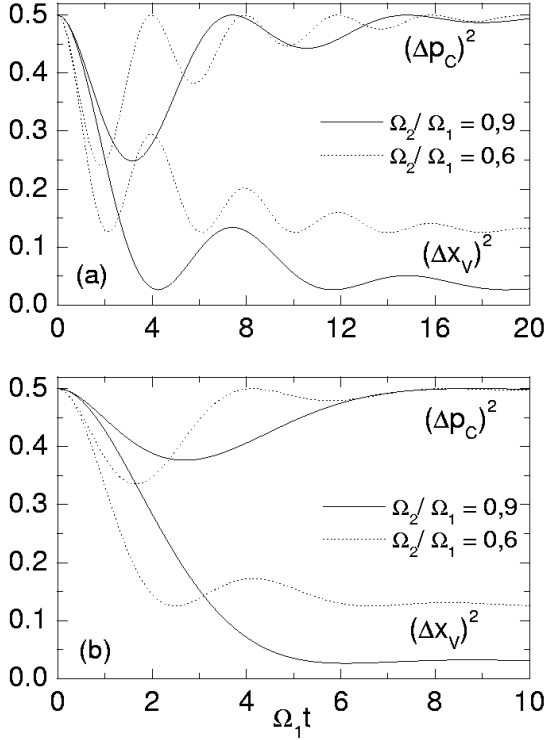}}
\caption{\label{fig:Figura2} $\Delta  P_c(t)^2$ and $\Delta  X_v(t)^2$ as a function of the adimensional time $\Omega_1 t$ for $\Omega_2/\Omega_1$ equals to $0.6$ and $0.9$ and for $\gamma/\Omega_1$ equals to a)  $0.4$  and b) $1.$  The squeezing $e^{-2\bar \xi}$ is equal to $0.25$ and $0.05$ for $\Omega_2/\Omega_1$ equals to $0.6$ and $0.9,$ respectively.}
\end{figure}

In Fig. 2 we show $\Delta P_c (t)^2$ and $\Delta X_v (t)^2 $ for $\Omega_2/\Omega_1$ equals to $0.6$ and $0.9$ and for $\gamma/\Omega_1$ equals to $0.4$ and $1.$ The maximum values of the squeezing for the $\hat X_v$ quadrature are reached at times $\tau_n$ when the vibrational state is equal to the squeezed vacuum state. At these times $\bar n_v(\tau_n)=0$ and the uncertainty in the quadrature $X_v$ is given by 
\begin{equation}\label{maximum}
\Delta X_v (\tau_n)^2=\frac12 e^{-2\bar \xi}=\frac12 \frac{\Omega_1-\Omega_2}{\Omega_1+\Omega_2}\, .
\end{equation}
 and is equal to $0.25$ and $0.05$ when $\Omega_2/\Omega_1$ equals to $0.6$ and $0.9,$ respectively. The quasi periodicity shown in the values of the uncertainties of the quadratures is a reflection of the fact that $f(t)$ and $g(t)$ are periodic functions attenuated by the exponential $e^{-\gamma t/4}.$ The values for $\gamma / \Omega_1 $ in Fig.~\ref{Figura2} are not far from those used in recent experiments~\cite{pachos,blatt}. For example, if we take $g_c/2\pi  = 6-7$MHz, $\gamma = (0.02-0.05)g_c$ and $g_1/ 2\pi\approx 10$MHz, $\Delta\approx 5 g_1$ and $\eta_c = 0.2$ we have $\gamma / \Omega_1 \approx 0.4-1.$

When $\Omega_2^2 <\Omega_1^2 \leq \Omega_2^2 + \gamma^2 /16,$ $\Lambda=i\vert\Lambda\vert$ is imaginary. In this case, as $t\rightarrow \infty, $  the functions $\cos(\Lambda t)$ and $\sin(\Lambda t)$ diverge. However the functions $f(t)$  and $g(t), $ given in Eq.~(\ref{ftgt}),  do converge to zero and the system still reaches the steady state given in Eq.~(\ref{steady}), although it does not present revivals. 

It is expected that recurrences of the whole system will occur when dissipation is absent. In fact, one should be able to diagonalize the quadratic Hamiltonian given by Eq.~(\ref{Veff}) through a Bogoliubov transformation and show that this system is equivalent to two independent harmonic oscillators and therefore revivals will occur for any initial condition. The remarkable result that we have shown above is that recurrences in {\it each individual system} occur when dissipation is present. 
For comparison we give below the time evolution of the system  when dissipation is absent and the initial state is the product of the two coherent states  $\vert \psi(0)\rangle=\hat D_c(\alpha)\vert 0\rangle_c\hat D_v(\beta)\vert 0\rangle_v:$ 
\begin{eqnarray}\label{gamma=0}
\vert \psi(t)\rangle 
&=&\hat D_c(u_0(t))\, \hat S_{c}(- \xi_0(t))\hat D_v(v_0(t))\, \hat S_{v}( \xi_0(t))\nonumber\\
&\times& \sum_{k=0}^{\infty} \frac{\bar n_0(t)^{k/2}}{(\bar n_0(t) + 1)^{(k+1)/2}}  
 \vert k \rangle_{c} \hskip 0.1cm  \, \vert k \rangle_{v}  \, ,
\end{eqnarray}
where 
\begin{eqnarray}\label{nexigamma=0}
&&\bar n_0(t) = \frac12\bigg( \sqrt{1+(\Omega_2^2/\Lambda_0^2)\sin^2(2\Lambda_0 t)}\, -1\bigg)  \, , \nonumber \\
&&\xi_0(t) = \frac14 \ln\bigg[\frac{(\Omega_1+\Omega_2)  (\Omega_1-\Omega_2 \, \cos (2\Lambda_0 t))}  
{(\Omega_1-\Omega_2)(\Omega_1+\Omega_2 \, \cos(2\Lambda_0 t))}\bigg]\, ,\nonumber\\
&& u_0(t)= \alpha\cos(\Lambda_0 t)+\bar\alpha\sin(\Lambda_0 t) \, ,\nonumber\\
&& v_0(t)=\beta\cos(\Lambda_0 t)+\bar\beta\sin(\Lambda_0 t)\, .
\end{eqnarray}
where $\Lambda_0^2=\Omega_1^2-\Omega_2^2,$ and
\begin{eqnarray}
\bar\alpha=(\beta^*\Omega_2+\beta\Omega_1)/\Lambda_0 \, , \nonumber\\
\bar\beta=(\alpha^*\Omega_2-\alpha\Omega_1)/\Lambda_0.
\end{eqnarray}

When $\Lambda_0 ^2> 0 ,$  $\vert \psi(t)\rangle$ is a periodic function of $t$ with a period $2\pi/\Lambda_0.$ When $t_m=m\pi/(2\Lambda_0), m=0,1,2,\dots, $ $\bar n_0(t_m)=0$ and the state of the motion and the state of the cavity field disentangle and are given by:
\begin{eqnarray}
\vert \psi_1\rangle &=&\hat D_c(\alpha)\vert 0\rangle_c\hat D_v(\beta)\vert 0\rangle_v\, , \nonumber\\
\vert \psi_2\rangle &= & \hat D_c(\bar\alpha)\, \hat S_{c}(- \bar \xi) \, \vert 0 \rangle_{c} \hskip 0.1cm \hat D_v(\bar\beta)\,\hat S_{v}( \bar \xi) \, \vert 0 \rangle_{v} \, ,\nonumber\\
\vert \psi_3\rangle &=&  \hat D_c(-\alpha)\vert 0\rangle_c\hat D_v(-\beta)\vert 0\rangle_v \, ,\nonumber\\
\vert \psi_4\rangle &=& \hat D_c(-\bar\alpha)\, \hat S_{c}(- \bar \xi) \, \vert 0 \rangle_{c} \hskip 0.1cm \hat D_v(-\bar\beta)\,\hat S_{v}( \bar \xi) \, \vert 0 \rangle_{v} \, ,
\end{eqnarray}
where $\bar\xi$ is given by Eq.~(\ref{xi}).

\subsection{\bf $\Omega_1^2 \le \Omega_2^2$}

When $\Omega_1^2 \le\Omega_2^2,$ the analytical results obtained in Eqs.(\ref{eqm}-\ref{Q00nxi})   are still valid, but the total system does not reach a steady state since it gains more energy than it is able to dissipate.
When $\Omega_1^2 < \Omega_2^2,$ $\xi_c(t)$ and $\xi_v(t)$ converge, as $t \rightarrow \infty,$ to the finite value given in  Eq.~(\ref{xi}). However $\bar n_c(t),$ $\bar n_v(t)$ and $f(t)g(t)$ increase monotonically, so that the ion-cavity system does not reach a steady state and remains always correlated. All uncertainties increase monotonically with $\Omega_2/\Omega_1.$

When $\Omega_1 = \Omega_2,$ the  system still remains  correlated all the time, since $f(t)g(t)$ never vanishes. Also,  both, the cavity field and the ion motion, do not  present either revival or squeezing. The uncertainties in the quadratures $\Delta P_c(t)^2$ and $\Delta X_v(t)^2$ are equal to $1/2$ during all the time, as we can see from Eq.~(\ref{DXDPfg}), while the other uncertainties are given by 
\begin{eqnarray}
\Delta X_c(t)^2&=& \frac12+\frac{8\Omega^2}{\gamma^2}(1-e^{-\gamma t/2})^2\, , \nonumber  \\
\Delta P_v(t)^2&=& \frac12+\frac{16\Omega^2}{\gamma^2}(\gamma t/2 +e^{-\gamma t/2}-1)^2 \, .
\end{eqnarray}
Notice that Eq.~(\ref{xi}) is no longer valid for $\Omega_2/\Omega_1=1,$ as we may not interchange the limits $\Omega_2\rightarrow\Omega_1$ and $t\rightarrow\infty$ in Eq~(\ref{zetaxin}).

\begin{figure}
\resizebox{!}{5cm}{\includegraphics{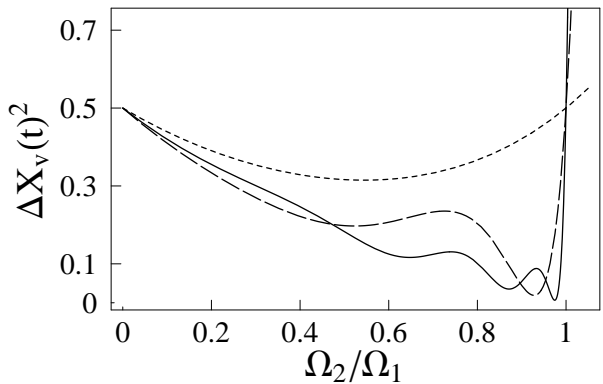}}
\caption{\label{fig:figura3} $\Delta X_v(t)^2$ as a function of $\Omega_2/\Omega_1$ for $\Omega_1 t$ 
equals to $1$ (dotted line), $5$ (dashed line) and $10$ (solid line). $\gamma/\Omega_1=0.4.$}
\end{figure}

In Fig.~\ref{figura3} we show the uncertainty  $\Delta X_v(t)^2$ as a function of $\Omega_2/\Omega_1$ for several times and for a fixed value of $\gamma=0.4\Omega_1.$ Notice that all curves  cross at $\Omega_2/\Omega_1=1, $ when $\Delta X_v(t)^2=1/2,$ and increase very fast for $\Omega_2/\Omega_1>1.$

\section{Conclusions}

We were able to provide an analytical solution for the total density operator of the system of a two-level ion trapped inside a quasi-resonant cavity subjected to the action of two laser fields with frequencies $\omega_c\pm\nu,$ when the initial state is a general coherent state.  The two external lasers  generate an  effective interaction that corresponds to: a) a parametric excitation of the vibrational mode and the cavity field (term proportional to $\Omega_2$) and b) an exchange of energy between them (term proportional to $\Omega_1$). When $\Omega_1^2 > \Omega_2^2$ the cavity mode and the ion vibrational motion are both always squeezed and the reduced density operator of each mode corresponds to a squeezed thermal state. 
The maximum value of the squeezing in the vibrational mode occurs with a period $\pi/\Lambda.$ When the initial state of the cavity field is the vacuum, the reduced density operator of the vibrational motion has ``revivals" of the squeezed vacuum state given in Eq.~(\ref{rhov}). The revivals in the motion state occur when the uncertainty in one of its quadratures reaches the minimum value given. The cavity field  returns to the initial state  periodically.  At  times when revivals occur the cavity and the motion  disentangle. 

When $\Omega_1^2 \le \Omega_2^2$ the system does not reach a steady state and does not present either revivals or squeezing.

The analytical solution for the total density operator may be useful in problems involving the interaction between two harmonic oscillators in the case that one of them is connected to a reservoir.

\acknowledgments

This work was supported by the Brazilian agencies: CNPq, FAPERJ and PRONEX. 

\begin{appendix}
\section{Solution of the master equation}

In Refs.~\cite{rangel,rangel2} the authors have obtained a solution for the master equation (\ref{eqm}) in the case $\Omega_2 = 0,$ by using ladder operators and an expansion in the eigenstates of the Liouvillian superoperator. We were not able  to find a simple similar solution in the case $\Omega_2 \neq 0$ using that method. Here we present the  solution of Eq.~(\ref{eqm}) using a different approach.
 
We are interested in finding the solution of the master equation (\ref{eqm}) when the initial state is the vacuum, \linebreak $\hat \rho_T(0)= \vert 0 \rangle_{c} \, _{c}\langle 0 \vert \vert 0 \rangle_{v} \, _{v}\langle 0 \vert .$ It may be written formally as follows: 
\begin{equation}\label{formal}
\hat \rho_T(t) = e^{{\cal K}t} \vert 0 \rangle_{c} \, _{c}\langle 0 \vert \vert 0 \rangle_{v} \, _{v}\langle 0 \vert   \, ,
\end{equation}
where ${\cal K}$ is the Liouvillian superoperator
\begin{equation}\label{K.}
{\cal K} = \frac{1  }{ i \hbar} (\hat V_{\rm eff}. - .\hat V_{\rm eff}) + \frac{\gamma }{ 2} (2 \hat a..\hat a^{\dagger} - \hat a^{\dagger} \hat a. - .\hat a^{\dagger} \hat a) \, ,
\end{equation}
with $\hat V_{\rm eff}$ being the effective Hamiltonian given in Eq.~(\ref{Veff}).

We have adopted the notation $\hat A.$ $(.\hat A)$ for superoperators that represent the simple action of an operator $\hat A$ to the left (right) on the target operator, $\hat \rho,$ i.e., $\hat A. \hat \rho:= \hat A \hat \rho$ $(.\hat A \hat \rho := \hat \rho \hat A).$ 

It will be convenient to define the following superoperators
\begin{eqnarray}\label{superoperators}
{\cal M}_+^c = \hat a^{\dagger}. - .\hat a^{\dagger} \quad &,& \quad {\cal N}_+^c = .\hat a - \hat a. \, , \nonumber \\
{\cal M}_-^c = \hat a.  \quad &,& \quad {\cal N}_-^c = .\hat a^{\dagger} \, , \nonumber \\
{\cal M}_+^v = \hat b^{\dagger}. - .\hat b^{\dagger} \quad &,& \quad {\cal N}_+^v = .\hat b - \hat b. \, , \nonumber \\
{\cal M}_-^v = \hat b.  \quad &,& \quad {\cal N}_-^v = .\hat b^{\dagger} \, ,
\end{eqnarray}
which obey the commutation relations
\begin{eqnarray}\label{}
[{\cal M}_-^c, {\cal M}_+^c] = 1 \quad &,& \quad  [{\cal N}_-^c , {\cal N}_+^c]=1 \, , \nonumber \\
\lbrack {\cal M}_-^v, {\cal M}_+^v \rbrack = 1 \quad &,& \quad  \lbrack {\cal N}_-^v , {\cal N}_+^v \rbrack =1 \, ,
\end{eqnarray}
while the remaining relations are null. The Liouvillian ${\cal K}$ may then be written as 
\begin{eqnarray}\label{Ksuper}
{\cal K} = {\cal K}^{\prime} + \Omega_2 ({\cal M}_+^c {\cal M}_+^v + {\cal N}_+^c {\cal N}_+^v)\, ,
\end{eqnarray}
where
\begin{eqnarray}\label{}
{\cal K}^{\prime} &=& - \gamma / 2 ({\cal M}_+^c {\cal M}_-^c + {\cal N}_+^c {\cal N}_-^c) \nonumber \\
&+& \Omega_1 ({\cal M}_+^c {\cal M}_-^v - {\cal M}_-^c {\cal M}_+^v) \nonumber \\ 
&+& \Omega_1 ( {\cal N}_+^c {\cal N}_-^v - {\cal N}_-^c {\cal N}_+^v) \nonumber \\
&+& \Omega_2 ({\cal M}_+^c {\cal N}_-^v + {\cal N}_-^c {\cal M}_+^v) \nonumber \\
&+& \Omega_2 ( {\cal N}_+^c {\cal M}_-^v + {\cal M}_-^c {\cal N}_+^v)\, .
\end{eqnarray}
From Eq.~(\ref{superoperators}) it is easy to see that 
\begin{equation}\label{M-00}
{\cal M}_-^{\sigma} \vert 0 \rangle_{\sigma} \, _{\sigma}\langle 0 \vert = 0  \quad , \quad {\cal N}_-^{\sigma} \vert 0 \rangle_{\sigma} \, _{\sigma}\langle 0 \vert = 0 \, ,
\end{equation}
where  $\sigma = c,v.$ Thus the superoperator ${\cal K'}$ annihilates the vacuum: 
\begin{equation}\label{Kprime0000}
{\cal K}^{\prime} \vert 0 \rangle_{c} \, _{c}\langle 0 \vert \vert 0 \rangle_{v} \, _{v}\langle 0 \vert = 0 \, .
\end{equation}
It is possible to relate ${\cal K}$ and ${\cal K}^{\prime}$ through the transformation
\begin{equation}\label{transf}
{\cal K} = e^{{\cal U}} {\cal K}^{\prime} e^{-{\cal U}} \, ,
\end{equation}
where
\begin{equation}\label{U}
{\cal U} = \frac{-\Omega_1 \Omega_2}{2(\Omega_1^2 - \Omega_2^2)} ({\cal M}_+^{v \hskip 0.1cm 2} + {\cal N}_+^{v \hskip 0.1cm 2}) + \frac{\Omega_2^2}{\Omega_1^2 - \Omega_2^2}{\cal M}_+^v {\cal N}_+^v \, .
\end{equation}
By exponentiation of Eq.~(\ref{transf}) we may rewrite the time evolution superoperator $e^{{\cal K}t}$ as
\begin{equation}\label{expKt}
e^{{\cal K}t} = e^{{\cal U}} e^{{\cal K}^{\prime}t} e^{-{\cal U}}\, . 
\end{equation}
Therefore the density operator at time $t$ may be written as
\begin{equation}\label{rhotA}
\hat\rho_T(t)= e^{{\cal U}} e^{{\cal K}^{\prime}t} e^{-{\cal U}}e^{{-\cal K}^{\prime}t} \vert 0 \rangle_{c} \, _{c}\langle 0 \vert \vert 0 \rangle_{v} \, _{v}\langle 0 \vert \, ,
\end{equation}
since $e^{-{\cal K}^{\prime}t}$ does not affect the vacuum state. The superoperator $e^{{\cal K}^{\prime}t} e^{-{\cal U}}e^{{-\cal K}^{\prime}t}$ may be written as
\begin{equation}\label{expKtU}
e^{{\cal K}^{\prime}t} e^{-{\cal U}}e^{{-\cal K}^{\prime}t} = \exp[-e^{{\cal K}^{\prime}t} {\cal U}e^{-{\cal K}^{\prime}t}] \, .
\end{equation}
To calculate the term in brackets in the above equation, we will need the following commutation relations:
\begin{eqnarray}\label{comutadorK}
&&\pmatrix{[{\cal K}^{\prime}, {\cal M}_+^c] \cr [{\cal K}^{\prime},{\cal N}_+^c] \cr [{\cal K}^{\prime},{\cal M}_+^v] \cr [{\cal K}^{\prime},{\cal N}_+^v] \cr} 
= \pmatrix{-\gamma/2 & 0 & -\Omega_1 & \Omega_2 \cr 0 & -\gamma/2 & \Omega_2 & - \Omega_1 \cr \Omega_1 & \Omega_2 & 0 & 0 \cr \Omega_2 & \Omega_1 & 0 & 0}\nonumber \\
&&\times \pmatrix{{\cal M}_+^c \cr {\cal N}_+^c \cr {\cal M}_+^v \cr {\cal N}_+^v \cr} .
\end{eqnarray}
From Eq.~(\ref{comutadorK}) and using the Baker-Haulsdorff formula, we get
\begin{eqnarray}\label{BCH+}
&& e^{{\cal K}^{\prime}t}\pmatrix{{\cal M}_+^c \cr {\cal N}_+^c \cr {\cal M}_+^v \cr {\cal N}_+^v \cr} e^{-{\cal K}^{\prime}t} \nonumber \\
&&= \pmatrix{h(t) & 0 & - \displaystyle{\frac{\Omega_1}{\Omega_2}} g(t) & g(t) \cr 0 & h(t) & g(t) & -\displaystyle{\frac{\Omega_1}{\Omega_2}} g(t) \cr \displaystyle{\frac{\Omega_1}{\Omega_2}} g(t) & g(t) & f(t) & 0 \cr g(t) & \displaystyle{\frac{\Omega_1}{\Omega_2}} g(t) & 0 & f(t) \cr} \nonumber \\
&& \times \pmatrix{{\cal M}_+^c \cr {\cal N}_+^c \cr {\cal M}_+^v \cr {\cal N}_+^v \cr}\, ,
\end{eqnarray}
where the functions $f(t),$ $g(t)$ and $h(t)$ are given by
\begin{eqnarray}\label{ftgtht}
f(t) &=& \bigg(\cos \Lambda t + \frac{\gamma}{4 \Lambda} \sin \Lambda t \bigg)e^{-\gamma t/4} \, ,\nonumber \\
g(t)&=&\frac{\Omega_2}{ \Lambda} \sin \Lambda t e^{-\gamma t/4} \, ,\nonumber \\
h(t)&=&\bigg(\cos \Lambda t - \frac{\gamma}{4 \Lambda} \sin \Lambda t \bigg)e^{-\gamma t/4}\, ,
\end{eqnarray}
with $\Lambda^2 = \Omega_1^2 - \Omega_2^2 - \gamma^2/16.$ The $4 \times 4$ matrix in Eq.~(\ref{BCH+}) was obtained by exponentiation of the $4 \times 4$ matrix in {Eq.~(\ref{comutadorK})}, what can be easily done by noticing that the latter may be rewritten in the form $- \gamma/4 {\bf 1} + \Lambda {\bf A},$ where ${\bf 1}$ is the $4 \times 4$ identity and ${\bf A}$ is a matrix such that ${\bf A}^2 = - {\bf 1}.$

Using Eq.~(\ref{U}), Eq.~(\ref{BCH+}) and that $[{\cal U},e^{-{\cal K}t} {\cal U} e^{{\cal K}t}]=0,$ we may rewrite Eq.~(\ref{rhotA})
\begin{equation}\label{rhoWt}
\hat\rho_T(t)=e^{{\cal W}(t)} \vert 0 \rangle_{c} \, _{c}\langle 0 \vert \vert 0 \rangle_{v} \, _{v}\langle 0 \vert \, , 
\end{equation}
where
\begin{eqnarray}\label{Vt}
{\cal W}(t) &=& {\cal U}- e^{{\cal K}^{\prime}t}{\cal U}e^{-{\cal K}^{\prime}t} \nonumber \\
&=& {\cal V}_c(\mu_c(t),\nu_c(t)) + {\cal V}_v(\mu_v(t),\nu_v(t)) \nonumber \\
&+& \zeta (t) ({\cal M}_+^c {\cal M}_+^v + {\cal N}_+^c {\cal N}_+^v)\, ,
\end{eqnarray}
with
\begin{equation}
{\cal V}_{\sigma}(\mu, \nu) = \mu ( {\cal M}_+^{\sigma \hskip 0.1cm 2} + {\cal N}_+^{\sigma \hskip 0.1cm 2} )/2 + \nu {\cal M}_+^{\sigma}{\cal N}_+^{\sigma}\, .
\end{equation}
The functions $\zeta(t),$ $\mu_c(t),$ $\nu_c(t),$ $\mu_v(t),$ and $\nu_v(t),$  in the above equation are given by 
\begin{eqnarray}\label{zetaxin2}
\zeta(t) &=&  f(t)g(t) \, , \nonumber \\
\mu_c(t)&=&  \frac{\Omega_1}{\Omega_2} \hskip 0.1cm g(t)^2  \, , \nonumber \\   
\nu_c(t)&=&  g(t)^2  \, , \nonumber \\ 
\mu_v(t)&=& - \frac{\Omega_1 \Omega_2}{\Omega_1^2 - \Omega_2^2} \hskip 0.1cm (1 - f(t)^2)   \, , \nonumber \\ 
\nu_v(t)&=& \frac{\Omega_2^2}{\Omega_1^2 - \Omega_2^2} \hskip 0.1cm (1 - f(t)^2) \, .
\end{eqnarray}

All the superoperators appearing in the definition of ${\cal W}(t)$ commute. Therefore we may write  
\begin{eqnarray}\label{expVt}
e^{{\cal W}(t)} &=& \exp[\, \zeta(t){\cal M}_+^c {\cal M}_+^v \, ] \,  \exp[\, \zeta(t){\cal N}_+^c {\cal N}_+^v)\, ] \nonumber \\
&\times& \exp[\, {\cal V}_c(\mu_c(t),\nu_c(t)) \, ] \nonumber \\
&\times& \exp[\, {\cal V}_v(\mu_v(t),\nu_v(t)) \, ]\, .
\end{eqnarray}
By substituting Eq.~(\ref{expVt}) into Eq.~(\ref{rhoWt}), we obtain 
\begin{eqnarray}
\hat \rho_T(t) &=& \sum_{m=0}^{\infty} \sum_{n=0}^{\infty} \zeta(t)^{m+n} \nonumber \\
&\times& \hat Q^{m,n}_{\hskip 0.1cm c}(\mu_c(t), \nu_c(t)) \, \hat Q^{m,n}_{\hskip 0.1cm v}(\mu_v(t), \nu_v(t))  \, , 
\end{eqnarray}
where 
\begin{equation}\label{defQmn}
\hat Q^{m,n}_{\hskip 0.1cm \sigma}(\mu,\nu) = \frac{{\cal N}_+^{\sigma \hskip 0.1cm n}}{\sqrt{n!}}\frac{{\cal M}_+^{\sigma \hskip 0.1cm m}}{\sqrt{m!}} \hat Q^{0,0}_{\hskip 0.1cm \sigma}(\mu,\nu)\, ,
\end{equation}
with
\begin{equation}\label{defQ00A}
\hat Q^{0,0}_{\hskip 0.1cm \sigma}(\mu,\nu) = e^{{\cal V}_{\sigma}(\mu,\nu)} \vert 0 \rangle_{\sigma} \, _{\sigma}\langle 0 \vert \, .
\end{equation}
Due the cyclic property of the trace we have the identities
\begin{equation}\label{trM+N+}
{\rm tr}_{\sigma}({\cal M}_+^{\sigma} \hat\rho) = 0 \quad , \quad {\rm tr}_{\sigma}({\cal N}_+^{\sigma} \hat\rho) = 0 \quad , \quad \forall \hat\rho \, .
\end{equation}  
Then we easily may conclude that
\begin{equation}\label{trQmnA}
{\rm tr}_{\sigma} \, \hat Q^{m,n}_{\sigma}(\mu,\nu) = \delta_{m,0} \delta_{n,0}\, .
\end{equation}

Below we give a useful expression for the operators $\hat Q^{0,0}_{\hskip 0.1cm \sigma}(\mu,\nu),$ which will allows us to write the reduced density operator $\hat \rho_\sigma(t)$ as a product of squeezing operators times an operator that describes a Planck's thermal distribution. In order to obtain this result we will start by defining three new superoperators:  
\begin{eqnarray}\label{ABC}
{\cal A}_{\sigma} &=& -({\cal M}_+^{\sigma \hskip 0.1cm 2} + {\cal N}_+^{\sigma \hskip 0.1cm 2})/2 - {\cal M}_+^{\sigma} {\cal N}_-^{\sigma} - {\cal M}_-^{\sigma} {\cal N}_+^{\sigma} \, , \nonumber \\
{\cal B}_{\sigma} &=& {\cal M}_+^{\sigma} {\cal N}_+^{\sigma} \, , \nonumber \\
{\cal C}_{\sigma} &=& {\cal M}_+^{\sigma} {\cal N}_-^{\sigma} + {\cal M}_-^{\sigma} {\cal N}_+^{\sigma} \, ,
\end{eqnarray}
which form the closed algebra:
\begin{eqnarray}
[{\cal A}_{\sigma}, {\cal B}_{\sigma}] &=& 2{\cal A}_{\sigma} + 2{\cal C}_{\sigma} \, , \nonumber \\
\lbrack {\cal A}_{\sigma},{\cal C}_{\sigma}\rbrack &=&  2 {\cal B}_{\sigma} \, , \nonumber \\
\lbrack {\cal B}_{\sigma},{\cal C}_{\sigma}\rbrack &=& 2{\cal A}_{\sigma} + 2{\cal C}_{\sigma} \, .
\end{eqnarray}
The superoperator ${\cal V}_{\sigma}(\mu,\nu)$ may be expressed in terms of these superoperators and, due to the fact that they form a Lie algebra, we may decompose $e^{{\cal V}_{\sigma}(\mu,\nu)}$ into a product of simpler exponentials,
\begin{eqnarray}
e^{{\cal V}_{\sigma}(\mu,\nu)} &=& e^{-\mu{\cal A}_{\sigma} + \nu{\cal B}_{\sigma} - \mu{\cal C}_{\sigma}} \nonumber \\
&=& e^{\xi{\cal A}_{\sigma}} e^{\bar n{\cal B}_{\sigma}} e^{\xi{\cal C}_{\sigma}} \, ,
\end{eqnarray}
where
\begin{eqnarray}\label{}
\bar n &=& - 1/2 + \sqrt{( \nu + 1/2)^2 - \mu^2}  \, ,\nonumber \\
\xi &=& \frac{1}{4} \hskip 0.1cm {\rm ln} \bigg( \frac{ \nu + 1/2 - \mu}{\nu + 1/2 + \mu}\bigg) \, .
\end{eqnarray}
From Eq.~(\ref{M-00}) and Eq.~(\ref{ABC}) we see immediately that 
\begin{equation}
e^{\xi{\cal C}_{\sigma}} \vert 0 \rangle_{\sigma} \, _{\sigma}\langle 0 \vert = \vert 0 \rangle_{\sigma} \, _{\sigma}\langle 0 \vert \, . 
\end{equation}
Also the operator
\begin{equation}\label{R00n}
\hat R^{0,0}_{\sigma}(\bar n) := e^{\bar n_{\sigma}{\cal B}_{\sigma}} \vert 0 \rangle_{\sigma} \, _{\sigma}\langle 0 \vert 
\end{equation}
may be calculated by noticing that
\begin{eqnarray}\label{eMe}
&&e^{\bar n_{\sigma}{\cal B}_{\sigma}} \, {\cal M}_-^{\sigma} \, e^{-\bar n_{\sigma}{\cal B}_{\sigma}} \, e^{\bar n_{\sigma}{\cal B}_{\sigma}} \vert 0 \rangle_{\sigma} \, _{\sigma}\langle 0 \vert = 0\, , \nonumber \\
&&e^{\bar n_{\sigma}{\cal B}_{\sigma}} \, {\cal N}_-^{\sigma} \, e^{-\bar n_{\sigma}{\cal B}_{\sigma}} \, e^{\bar n_{\sigma}{\cal B}_{\sigma}} \vert 0 \rangle_{\sigma} \, _{\sigma}\langle 0 \vert = 0 \, ,
\end{eqnarray}
and that 
\begin{eqnarray}\label{}
&&e^{\bar n_{\sigma}{\cal B}_{\sigma}} \, {\cal M}_-^{\sigma} \, e^{-\bar n_{\sigma}{\cal B}_{\sigma}} = {\cal M}_-^{\sigma} - \bar n {\cal N}_+^{\sigma} \, , \nonumber \\
&&e^{\bar n_{\sigma}{\cal B}_{\sigma}} \, {\cal N}_-^{\sigma} \, e^{-\bar n_{\sigma}{\cal B}_{\sigma}} = {\cal N}_-^{\sigma} - \bar n {\cal M}_+^{\sigma} \, .
\end{eqnarray}
Then $\hat R^{0,0}_{\sigma}(\bar n)$ satisfies 
\begin{eqnarray}\label{M-N+R00}
&&({\cal M}_-^{\sigma} - \bar n {\cal N}_+^{\sigma})\hat R^{0,0}_{\sigma}(\bar n) = 0 \, , \nonumber \\
&&({\cal N}_-^{\sigma} - \bar n {\cal M}_+^{\sigma})\hat R^{0,0}_{\sigma}(\bar n) = 0 \, .
\end{eqnarray}
Using the identities (\ref{trM+N+}) and the definition (\ref{R00n}) we easily get 
\begin{equation}\label{trR00n}
{\rm tr}_{\sigma}\hat R^{0,0}_{\sigma}(\bar n) = 1 \, . 
\end{equation}
The solution of Eq.~(\ref{M-N+R00}) under the normalization condition given in Eq.~(\ref{trR00n}) is a thermal state:
\begin{equation}\label{R00nA}
\hat R^{0,0}_{\sigma}(\bar n) = \sum_{k=0}^{\infty} \frac{\bar n^k}{(\bar n + 1)^{k+1}}\vert k \rangle_{\sigma} \, _{\sigma}\langle k \vert \, .
\end{equation}
Thus the operator $\hat Q^{0,0}_{\sigma}(\bar n,\xi)$ may be rewritten as 
\begin{equation}\label{Q00A}
\hat Q^{0,0}_{\sigma}(\bar n,\xi) = e^{\xi{\cal A}_{\sigma}}\hat R^{0,0}_{\sigma}(\bar n)  \, .
\end{equation}
It is possible to see that 
\begin{equation}\label{expA}
e^{\xi {\cal A}_{\sigma}} = \hat S_{\sigma}(\xi)..\hat S_{\sigma}(\xi)^{\dagger}\, ,
\end{equation}
where $\hat S_{\sigma}(\xi)$ are the squeezing operators 
\begin{eqnarray}\label{SA}
\hat S_c(\xi) &=& e^{\xi/2 (\hat a^2 - \hat a^{\dagger \hskip 0.05cm 2})} \, , \nonumber \\ 
\hat S_v(\xi) &=& e^{\xi/2 (\hat b^2 - \hat b^{\dagger \hskip 0.05cm 2})} \, .
\end{eqnarray}
By using Eq.~(\ref{R00nA}), Eq.~(\ref{Q00A}) and Eq.~(\ref{expA}), we obtain 
\begin{eqnarray}\label{Q00nxiA}
&&\hat Q^{0,0}_{\hskip 0.1cm \sigma}(\bar n,\xi) \nonumber \\
&&= \sum_{k=0}^{\infty} \frac{\bar n^k}{(\bar n+1)^{k+1}} \hat S_{\sigma}(\xi) \, \vert k \rangle_{\sigma} \, _{\sigma}\langle k \vert \, \hat S_{\sigma}(\xi)^{\dagger} \, , 
\end{eqnarray}
which is the same as Eq.~(\ref{Q00nxi}) of Sec.~III.

By using Eq.~(\ref{Q00A}) and inserting identities $1 = e^{\xi {\cal A}_{\sigma}} e^{-\xi {\cal A}_{\sigma}}$ into the definition of $\hat Q^{m,n}_{\sigma}(\bar n,\xi),$ Eq.~(\ref{defQmn}), we obtain
\begin{eqnarray}\label{Qnxi}
&&\hat Q^{m,n}_{\sigma}(\bar n,\xi)= e^{\xi {\cal A}_{\sigma}} \, \frac{(e^{-\xi {\cal A}_{\sigma}}{\cal N}_+^{\sigma} \, e^{\xi {\cal A}_{\sigma}})^n}{\sqrt{n!}} \nonumber \\
&&  \times \frac{(e^{-\xi {\cal A}_{\sigma}}{\cal M}_+^{\sigma} \, e^{\xi {\cal A}_{\sigma}})^m}{\sqrt{m!}} \, \hat R^{0,0}_{\hskip 0.1cm \sigma}(\bar n) \, .
\end{eqnarray}
By substituting the equalities
\begin{eqnarray}
e^{-\xi {\cal A}_{\sigma}}{\cal M}_+^{\sigma} \, e^{\xi {\cal A}_{\sigma}} &=& \cosh \xi \, {\cal M}_+^{\sigma} + \sinh \xi \, {\cal N}_+^{\sigma}\, , \nonumber \\
e^{-\xi {\cal A}_{\sigma}}{\cal N}_+^{\sigma} \, e^{\xi {\cal A}_{\sigma}} &=& \sinh \xi \, {\cal M}_+^{\sigma} + \cosh \xi \, {\cal N}_+^{\sigma}\, ,
\end{eqnarray}
into Eq.~(\ref{Qnxi}) we obtain
\begin{eqnarray}\label{QmnnxiA}
&&\hat Q^{m,n}_{\hskip 0.1cm \sigma}(\bar n,\xi)  \nonumber \\
&& = \sum_{k=0}^{m+n} C^{m,n}_{\hskip 0.1cm k}(\xi) \, \hat S_{\sigma}(\xi) \, \hat R^{m+n-k,k}_{\hskip 0.2cm \sigma} (\bar n) \, \hat S_{\sigma}(\xi)^{\dagger} \, ,
\end{eqnarray}
where we have defined
\begin{equation}
\hat R^{m,n}_{\hskip 0.1cm \sigma}(\bar n) = \frac{{\cal N}_+^{\sigma \hskip 0.1cm n}}{\sqrt{n!}} \frac{{\cal M}_+^{\sigma \hskip 0.1cm m}}{\sqrt{m!}} \hat R^{0,0}_{\hskip 0.1cm \sigma}(\bar n) 
\end{equation}
and 
\begin{eqnarray}\label{CmnA}
&&C^{m,n}_{\hskip 0.1cm k}(\xi) = \sqrt{\frac{(m+n-k)!k!}{m!n!}} \nonumber \\
&\times& \sum_{l={\rm max}(0,k-m)}^{{\rm min}(n,k)} \frac{m!}{(k-l)!(m-k+l)!}\frac{n!}{l!(n-l)!} \nonumber \\
&\times& (\cosh \xi)^{m-k+2l} (\sinh \xi)^{n+k-2l}\, .
\end{eqnarray}
Using Eq.~(\ref{R00nA}) and Eq.~(\ref{superoperators}), we obtain the expression of $\hat R^{m,n}_{\hskip 0.1cm \sigma}(\bar n),$ for $m \geq n,$ in the Fock basis 
\begin{eqnarray}\label{RmnnA}
&&\hat R^{m,n}_{\hskip 0.1cm \sigma}(\bar n)  = \sum_{k = 0}^{\infty} \hskip 0.1cm \sqrt{\frac{n! k! }{ m! (k + m -n)!}} \hskip 0.1cm \frac{1 }{ (\bar n + 1)^{m+1}}  \nonumber \\
&\times& P_{\hskip 0.1cm \hskip 0.1cm m}^{\hskip 0.1cm k,k -n} \biggl(\displaystyle{\frac{\bar n }{ \bar n +1}}\biggr) \vert k + m -n \rangle_{\sigma} \, _{\sigma}\langle k\vert \, , 
\end{eqnarray} 
where
\begin{equation}
P_{\hskip 0.1cm \hskip 0.1cm m}^{\hskip 0.1cm k,l}(x) = \sum_{j={\rm max} (0,l)}^{k} (-1)^{j-l} \frac{(j + m)! }{ (j-l)! (k-j)!} \frac{x^j }{ j!}\, .
\end{equation}
The expression of $\hat R^{m,n}$ for $m \leq n$ can be easily obtained from the property $\hat R^{m,n}=( \hat R^{n,m})^{\dagger}.$ 

\end{appendix}

\end{document}